\documentclass[a4paper]{article}

\usepackage{icrc2013}
\usepackage[english]{babel}
\usepackage{hyperref}

\newcommand{\ctools}{\mbox{\em ctools}}
\def\deg{\ensuremath{^\circ}}

\title{Towards a common analysis framework for gamma-ray astronomy}

\shorttitle{Towards a common analysis framework for gamma-ray astronomy}

\authors{
J\"urgen Kn\"odlseder$^{1,2}$, 
Michael Mayer$^{3,4}$,
Christoph Deil$^5$,
Anneli Schulz$^3$,
Marie-H\'el\`ene Grondin$^{1,2}$,
Pierrick Martin$^{1,2}$,
and Sylvie Brau-Nogu\'e$^{1,2}$
for the CTA Consortium.
}

\afiliations{
$^1$ Universit\'e de Toulouse; UPS-OMP; IRAP; Toulouse, France\\
$^2$ CNRS; IRAP; 9 Av. colonel Roche, BP 44346, F-31028 Toulouse Cedex 4, France\\
$^3$ DESY, Platanenallee 6, D-15738 Zeuthen, Germany \\
$^3$ Universit\"at Potsdam, Am Neuen Palais 10, D-14469 Potsdam, Germany \\
$^4$ Max-Planck-Institut f\"ur Kernphysik, P.O. Box 103980, D-69029 Heidelberg, Germany
}

\email{jknodlseder@irap.omp.eu}

\abstract{
Thanks to the success of current gamma-ray telescopes (Fermi, H.E.S.S., MAGIC, VERITAS), and 
in view of the prospects of planned observatories such as the Cherenkov Telescope Array (CTA) 
or the High-Altitude Water Cherenkov Observatory (HAWC), gamma-ray astronomy is becoming 
an integral part of modern astrophysical research. 
Analysis today relies on a large diversity of tools and software frameworks that were specifically 
and independently developed for each instrument. With the aim of unifying the analysis of gamma-ray 
data, we are currently developing GammaLib (\url{http://sourceforge.net/projects/gammalib}), a C++ library
interfaced to Python that provides a framework for an instrument independent analysis of gamma-ray 
data.
On top of GammaLib we have created \ctools\ (\url{http://cta.irap.omp.eu/ctools}), a set of analysis executables 
that is being developed as one of the prototypes for the CTA high-level science analysis framework, but 
which is equally suited for the analysis of gamma-ray data from the existing Fermi-LAT telescope and 
current Cherenkov telescope arrays. 
In particular, \ctools\ and GammaLib provide the novel opportunity of a simultaneous multi-instrument 
analysis. 
We present the status of the software development, and illustrate its capabilities with a spectral analysis
of the Crab nebula emission over seven decades in energy (1 MeV to 10 TeV) using multi-instrument
(COMPTEL, Fermi-LAT, H.E.S.S.) gamma-ray observations as well as a simulation of a CTA observation
of the supernova remnant RX~J1713.7$-$3946.
}


\begin{document}
\maketitle

\section{Introduction}

The field of gamma-ray astronomy has experienced a spectacular progress during the last
decade, thanks to significant improvements in the performance of ground-based and 
space-based gamma-ray telescopes \cite{holder2012, michelson2010}.
Ground-based imaging atmospheric Cherenkov telescopes (IACTs), such as H.E.S.S., VERITAS and 
MAGIC, have detected more than 100 sources of gamma rays between a few tens of GeV up 
to $\sim10$ TeV, unveiling cosmic particle acceleration in Galactic (PWN, SNR, gamma-ray
binaries) and extragalactic objects (starburst galaxies and AGN).
The Fermi Gamma-Ray Space Telescope, which explores the sky since 2008 at energies
from 30 MeV to 300 GeV, has detected so far almost 2000 sources of gamma rays
\cite{nolan2012}, comprising large populations of pulsars and AGN in addition to PWN, SNR, 
gamma-ray binaries, globular clusters, normal galaxies, starburst galaxies and radio galaxies.

A comparable amount of sources is expected to be detected by the planned Cherenkov Telescope 
Array (CTA) \cite{acharya2013}.
One of the major challenges of CTA is that it will be the first IACT being operated as an 
open observatory.
CTA will accept observing proposals from interested scientists and provide tools and support for 
data analysis that will be compliant with existing standards, such as for example the
FITS data format \cite{pence2010} or HEASARCs FTOOLS \cite{pence1993}.
This will considerably differ from practices used for data analysis of existing IACTs, where 
dedicated integrated analysis frameworks and custom data formats that are proprietary to
the respective collaborations are the rule.

In this paper we present an open source framework for scientific analysis of astronomical 
gamma-ray data that implements methods (e.g. maximum likelihood fitting,
source and background modeling) and standards (e.g. FITS data format, IRAF parameter
interface \cite{valdes1992}) that are readily employed for analyses of space-based 
high-energy telescope data and makes them available for analysis of IACT data. 
The core of this framework is the GammaLib, an open source C++ library that implements all code 
required for high-level science analysis of astronomical gamma-ray data in an 
instrument-independent way.
This allows for the analysis of data produced by a variety of different space-based and
ground-based telescopes, and enables gathering of data from different instruments for a combined
and coherent multi-instrument analysis.

On top of GammaLib we have created \ctools, a set of analysis executables that is largely inspired
from HEASARCs FTOOLS \cite{pence1993} and Fermi's Science Tools, allowing the assembly of 
modular workflows for IACT data analyses.
The \ctools\ have been successfully tested on existing IACT data from H.E.S.S. and MAGIC, as well
as on simulated data from CTA \cite{dubus2013}.
We propose the \ctools\ to be used as the public Science Tools for high-level science analysis of CTA 
data.

\section{GammaLib}

\begin{figure}[t]
  \centering
  \includegraphics[width=0.4\textwidth]{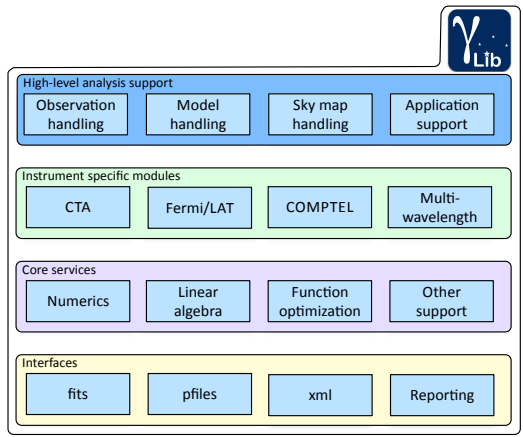}
  \caption{Organization of the GammaLib library.}
  \label{fig:gammalib}
\end{figure}

GammaLib provides a unified framework for the high-level analysis of astronomical gamma-ray data\break
(\url{http://sourceforge.net/projects/gammalib}).
The GammaLib development has been initiated in 2006 at the Institut de Recherche en Astrophysique
et Plan\'etologie (IRAP) in Toulouse (France), and counts today a growing number of developers from
various institutes and countries (\url{https://www.ohloh.net/p/GammaLib}).
Although GammaLib does not specifically focus on science analysis of IACT data, a good fraction of
the developments during the past 2 years have been motivated by the needs of the CTA project.

GammaLib is organized into four software layers that comprise several modules
(see figure \ref{fig:gammalib}).
The top layer provides support for instrument-independent data analysis, comprising information 
related to observations (event data, pointing and exposure information), instrument response 
functions, parametric source and background models used for model fitting, and
sky maps.
Also the support for creating FTOOLS applications is part of this layer.
All instrument specific interfaces are implemented in the next layer, with one module per gamma-ray
telescope.
So far, GammaLib provides support for analysis of data from the 
COMPTEL telescope,
the Fermi-LAT telescope, and
CTA (including also support for others IACTs).
Usage of multi-wavelength data in form of spectral energy distributions (SEDs) is also supported
through a dedicated module.
Core services related to numerical computations and function optimization are implemented in
the next layer.
Finally, an interface layer allows handling of data in FITS format and in XML format.
This layer also implements the IRAF parameter interface and reporting.

\section{{\em ctools}}

The way how GammaLib can be used is twofold:
GammaLib classes can be included and instantiated in a compiled C++ executable, or they can
be called directly from Python via a dedicated module.
Both methods have been used to implement the \ctools\ (\url{http://cta.irap.omp.eu/ctools}).

\ctools\ is a set of tools which each performs a single, well-defined analysis
step.
These steps comprise
event simulation ({\em ctobssim}),
event selection ({\em ctselect}),
event binning ({\em ctbin}), and
binned or unbinned maximum likelihood model fitting ({\em ctlike}).
The \ctools\ philosophy is very similar to the rational behind the FTOOLS \cite{pence1993}, which
are widely used in X-ray astronomy, and which also have inspired the science analysis frameworks
of INTEGRAL and Fermi.
\ctools\ operates on high-level CTA event lists, i.e. reconstructed events that have been
calibrated in energy and from which most of the particle background has been removed on
basis of air Cherenkov shower image characteristics (IACT event reconstruction and background
discrimination is thus not part of \ctools).

Each of the tools is created as a derived class of the GammaLib class {\tt GApplication}, which
provides a standard user interface and common functionalities and behavior to all of the tools.
In particular, tools that are implemented as Python scripts (dubbed {\em cscripts}) will show identical
interfaces and behavior as tools implemented as compiled C++ executables, making them
indistinguishable to the user.
Python scripts are mainly used for prototyping and in case that customizable tools are needed,
while C++ executables are used for production tools and tools where maximum computational speed
is critical.

All \ctools\ can be called from the command line using the IRAF parameter interface.
They can also be scripted from shell scripts, or they can be called directly from Python via a 
dedicated module.
Using \ctools\ from Python avoids the need for storing intermediate results on disk, as data can
be passed directly in memory from one tool to the other.
This enables the creation of purely in-memory analysis workflows for scientific analyses.

\section{Applications}

In the following sections we show several applications of \ctools\ (and thus also GammaLib)
that demonstrate the current capabilities and illustrate the potential for future usage
for CTA.

\subsection{Maximum likelihood model fitting}

As a first application we perform a maximum likelihood fit of a source model on top of a model
for the residual particle background to $\sim2$ hours of H.E.S.S. data of the Crab nebula.
This analysis method is close to the spatio-spectral fitting that is employed for the analysis of
Fermi-LAT data (and that is implement in the Fermi-LAT science tool {\em gtlike}), and differs substantially
from the conventional methods employed in VHE astronomy, which are mostly based on aperture 
photometry and background modeling from off regions. 

The data that we used have been provided by the H.E.S.S. collaboration to the CTA collaboration
in the context of the first CTA Data Challenge (CTA-1DC), and consist of 4 runs of $\sim28$ minutes
length taken with offsets of 0.5\deg\ and 1.5\deg\ from the source position.
The event data as well as the associated effective areas and point spread function have been
stored in FITS file format, no energy redistribution information is used for the analysis.

The Crab nebula has been modeled as a point source with a power-law energy spectrum.
The spatial distribution of the events for a point source has been described by a superposition of
three 2D Gaussian functions with energy-dependent widths and relative amplitudes.
The particle background has been modeled using a spatial model of the form
\begin{equation}
B(\theta) \propto \exp \left( -\frac{1}{2} \frac{\theta^4}{\sigma^2} \right)
\label{eq:background}
\end{equation}
where
$\theta$ designates the angle between the center of the camera and the reconstructed direction of 
the event, and
$\sigma$ is a width parameter.
The energy dependent count rate of the particle background has been modeled using a piecewise 
power law, defined by the background rate at six energies spanning the analysis interval.
All events in the energy band $0.5-20$ TeV have been used for the analysis.
The free parameters of the analysis were the position of the source, the prefactor and the spectral
index of the power law, the width of the 2D Gaussian for the particle background, and the six background
rates.

We first performed a binned maximum likelihood analysis, for which all events for a given run
have been binned in $100\times100$ spatial pixels of size $0.02\deg\times0.02\deg$ around the 
Crab nebula (resulting in a squared analysis region of about $2\deg\times2\deg$ around the 
Crab nebula).
Ten logarithmically-spaced bins in energy have been adopted.
Instead of summing the events of the individual runs into a single counts cube, we created one
counts cube per run and performed the analysis by maximizing the sum of the likelihood of all
pixels in the 4 runs.

\begin{figure}[t]
  \centering
  \includegraphics[width=0.45\textwidth]{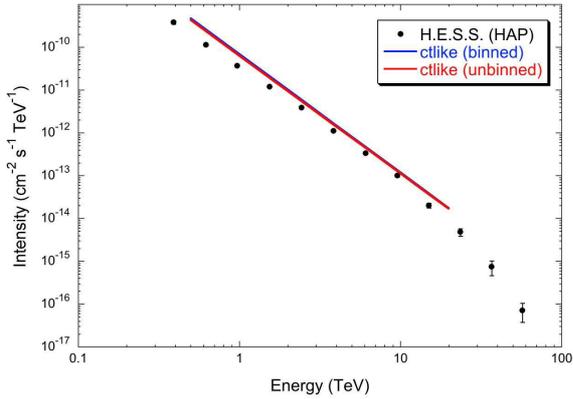}
  \caption{Comparison of {\em ctlike} fitted spectra to published H.E.S.S. results \cite{khelifi2008}.}
  \label{fig:maxlike}
\end{figure}

The binned analysis resulted in a maximum likelihood position of the Crab nebula that was offset
by less than $1$ arcmin from the nominal position.
The fitted power law spectral parameters are illustrated in figure \ref{fig:maxlike}, where we show
the best fitting power law (blue line) in comparison to the spectral points published by the H.E.S.S. 
using the HAP software framework \cite{khelifi2008}.
Our fitted spectrum is globally above the H.E.S.S. flux points, and further investigations are needed
to understand this offset.
Note, however, that the published H.E.S.S. analysis covers a different and substantially larger dataset
than the one used for our analysis, which could explain the apparent differences.

As next step we performed an unbinned maximum likelihood analysis, for which all events within
a radius of 1\deg\ of the Crab nebula have been used.
The analysis covers the same energy range as the binned analysis.
The maximum likelihood fitting results obtained with the unbinned analysis are extremely close to
those obtained with the binned analysis, demonstrating the consistency between both approaches
(note that binned and unbinned analyses do not use exactly the same events, as the binned
analysis is done on a squared $2\deg\times 2\deg$ analysis region while the unbinned analysis
is done using a circular region with $2\deg$ in diameter).
Figure \ref{fig:maxlike} shows also the power law obtained using the unbinned analysis (red  line).
Both power laws are almost indistinguishable, illustrating the consistency between both analysis
methods.

\subsection{Multi-instrument analysis}

\begin{figure}[t]
  \centering
  \includegraphics[width=0.45\textwidth]{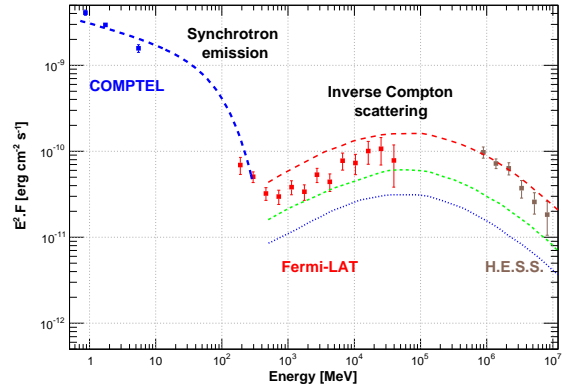}
  \caption{Gamma-ray SED of the Crab nebula derived using {\em ctlike}.}
  \label{fig:multispec}
\end{figure}

As next step we extend our analysis by adding observations of the Crab nebula performed
with the COMPTEL and the Fermi-LAT telescopes.
We then optimize all model parameters in a joint likelihood fit, where the likelihood function is
computed by summing over all events in all datasets.
The COMPTEL data have been retrieved from the HEASARC archive 
(\url{http://heasarc.gsfc.nasa.gov/docs/cgro/archive/})
and comprise about two weeks of continuous observations of the Crab nebula (viewing period 1.0).
Three binned event cubes have been included in the analysis, spanning the energy ranges
$0.75-1$, $1-3$ and $3-10$ MeV.
The Fermi-LAT data have been retrieved from the Fermi Science Support Center 
(FSSC: \url{http://fermi.gsfc.nasa.gov/ssc/}).
Pulse phases were assigned to the data using the Fermi plug-in provided by the LAT team and 
distributed with TEMPO2 and an ephemeris built from radio observations (available on the FSSC). 
Following the procedure described in \cite{Abdo2010}, photons from the unpulsed phase interval
(corresponding to 25\% of the total phase range) were selected, to avoid any contamination from the 
Crab Pulsar. 
This dataset will be considered in the following analysis.

Since the Crab nebula does not obey a single power law over such a wide energy range, we replaced
the single power law for the Crab nebula by a piecewise power law that we fitted over the 
$0.75$ MeV - $10$ TeV energy range.
The energies at which the intensities of the piecewise power law have been fitted had been selected
to cover evenly the energy ranges covered by the instruments.
In total, 22 intensities parameters have been fitted for the Crab nebula.
The fitted values of the intensity parameters are shown as flux points in figure \ref{fig:multispec}.
The fit of the synchrotron component using COMPTEL and Fermi-LAT data (dashed blue line) 
and the intensities predicted for the Inverse Compton scattering for 3 different magnetic fields 
(red solid line: 100 $\mu$G, green dashed line: 200 $\mu$G, blue dotted line: 300 $\mu$G) 
are overlaid for comparison. 
These curves are taken from \cite{Abdo2010}.

Fitting a piecewise power law to multi-instrument data with non-overlapping energy coverage
does not really warrant a joint analysis, as the intensity parameters are basically uncorrelated
between the instruments.
Separate analyses of the COMPTEL, Fermi-LAT and H.E.S.S. data would in fact have led to the 
same results.
The situation is different when parametric spectral models with few (eventually physical meaningful)
parameters are used to describe the data.
Then, the joint datasets will constrain the parameters covariantly, allowing in particular for a
coherent assessment of parameter uncertainties. 

\begin{figure}[t]
  \centering
  \includegraphics[width=0.45\textwidth]{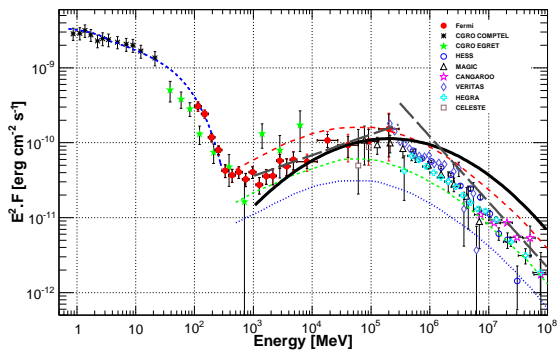}
  \caption{Broad-band fit using {\em ctlike}.}
  \label{fig:broadband}
\end{figure}

For illustration, we show in figure \ref{fig:broadband} the result of jointly fitting a log-parabola model
$I_{\gamma}(E) = k (E/E_{\rm p})^{-\alpha -\beta \log E/E_{\rm p}}$ to the Fermi-LAT
and H.E.S.S. data of the Crab nebula using the {\em ctlike} tool (thick black line).
For comparison, we also show the flux points for the Crab nebula from the literature that have
been derived from data of various instruments.
The dashed grey lines present power law fits to the Fermi-LAT data and the H.E.S.S. data alone.

\subsection{Extended emission}

The \ctools\ software is not only suited for spectral fitting, it also provides support for studying the
gamma-ray emission morphology.
A typical use case will be the determination of the radius and width of a supernova remnant
shell, such as the one observed for the first time from RX~J1713.7$-$3946 in VHE gamma-rays.

\begin{figure}[t]
  \centering
  \includegraphics[width=0.45\textwidth]{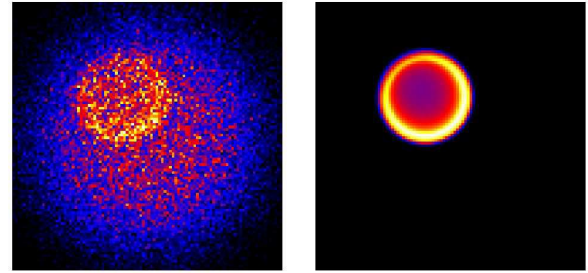}
  \caption{Simulation (left) and shell model (right) of RX~J1713.7$-$3946.}
  \label{fig:shell}
\end{figure}

To illustrate this use case, we simulated a CTA observation of RX~J1713.7$-$3946 with an
exposure of 5 hours using {\em ctobssim} for array configuration E \cite{bernlohr2013}.
The supernova remnant has been modeled as a shell with an apparent inner radius of 
$0.7\deg$ and an apparent thickness of $0.2\deg$.
A power law spectrum with index of $-2.19$ and an integral flux above 1 TeV of 
$1.46 \times 10^{-11}$ ph cm$^{-2}$ s$^{-1}$ has been assumed.
The particle background has been modeled using equation \ref{eq:background} with
$\sigma=3\deg$.
Figure \ref{fig:shell} shows in the left panel the counts map of this observation obtained using
{\em ctbin}.

We then used {\em ctlike} to fit the data using a shell model, where the shell position, inner shell 
radius and the shell thickness were left as free parameters.
Also the power law spectral index and integral flux were left free, as well as the normalization and
size of the particle background.
The model fit has been performed using an unbinned maximum likelihood analysis.
The right panel of figure \ref{fig:shell} shows the fitted model of RX~J1713.7$-$3946 that we
computed using {\em ctmodel}.
The asymmetric appearance of the model is due to the drop of the effective area when moving
away from the centre of the field of view.
The match between simulated data and fitted model is very satisfactory. 

\section{Conclusions}

We have developed a common analysis framework for gamma-ray astronomy data that can
potentially be used for the analysis of any type of event data.
So far, interfaces have been implemented to support the analysis of COMPTEL, Fermi-LAT
and IACT data.
We have shown several applications that illustrate the current capabilities of the framework,
including joint multi-instrument spectral analyses and morphology studies.
The basic building blocks of the framework are now implemented and tested;
future work will be dedicated to expand the support to additional gamma-ray telescopes, 
and to enrich the existing interfaces for more complex analyses.

We put here the emphasis on demonstrating the potential of our framework for the future CTA
observatory, allowing the scientific analysis of the observatory's data itself, and enabling the
joint analysis of CTA data with data from other instruments, such as those from the Fermi-LAT
telescope.

We still need to demonstrate that GammaLib and \ctools\ can cope with the complex particle
background that is encountered in VHE astronomy (so far, tests have only been done on the Crab
nebula, which is a bright point source for which the particle background modeling is less important).
This will be achieved by applying our tools to existing data from H.E.S.S. and the other
active VHE experiments.
Confronting the framework to real data will allow us to refine the particle background modeling
methods, and to demonstrate the validity of our approach.
We furthermore plan to implement also the conventional VHE analysis methods in GammaLib
and \ctools, enabling cross-checking with results obtained by the existing analysis chains.

We finally recall that GammaLib and \ctools\ are open source community tools.
The software can be freely downloaded from 
\url{http://sourceforge.net/projects/gammalib}
and 
\url{http://cta.irap.omp.eu/ctools}, and we invite everybody interested in using the tools to do so,
or even better, to join the development team for making the product even better.

\vspace*{0.5cm}
\footnotesize{{\bf Acknowledgment:}{
We gratefully acknowledge support from the agencies and organizations listed in this page: 
http://www.cta-observatory.org/?q=node/22}.
We also acknowledge the H.E.S.S. and MAGIC collaborations for releasing some data for CTA-1DC.}


\end{document}